\documentclass[amsmath,amssymb,prl,reprint]{revtex4-1}
\usepackage{graphicx}
\usepackage{dcolumn}
\usepackage{bm}
\usepackage{natbib}

\begin{document}

\title{On the accuracy of conductance quantization in spin-Hall insulators}
\author{S.K.~Konyzheva}
\affiliation{Institute of Solid State Physics, Russian Academy of
Sciences, 142432 Chernogolovka, Russian Federation}
\affiliation{Moscow Institute of Physics and Technology, Dolgoprudny, 141700 Russian Federation}
\author{E.S.~Tikhonov}
\affiliation{Institute of Solid State Physics, Russian Academy of
Sciences, 142432 Chernogolovka, Russian Federation}
\affiliation{Moscow Institute of Physics and Technology, Dolgoprudny, 141700 Russian Federation}
\author{V.S.~Khrapai}
\affiliation{Institute of Solid State Physics, Russian Academy of
Sciences, 142432 Chernogolovka, Russian Federation}
\affiliation{Moscow Institute of Physics and Technology, Dolgoprudny, 141700 Russian Federation}

\begin{abstract}
In contrast to the case of ordinary quantum Hall effect, the resistance of ballistic helical edge channels in typical quantum spin-Hall experiments is non-vanishing, additive and poorly quantized. Here we present a simple argument connecting this qualitative difference with a spin relaxation in the current/voltage leads in an experimentally relevant multi-terminal bar geometry. Both the finite lead resistance and the spin relaxation contribute to a non-vanishing four-terminal edge resistance, explaining poor quantization quality. We show that corrections to the four-terminal and two-terminal resistances in the limit of strong spin relaxation are opposite in sign, making a measurement of the spin relaxation resistance feasible, and estimate the magnitude of the effect in HgTe-based quantum wells.
\end{abstract}

\maketitle

Transverse conductance in an ordinary quantum Hall effect is quantized with metrological accuracy. In the effective Landauer-B\"uttiker description~\cite{Buettiker1988} this is interpreted as a conductance quantization of one-dimensional (1D) edge channels. Such edge channels are protected by chirality thereby their four-terminal resistance vanishes. By contrast, conventional 1D systems inevitably suffer from contact effects~\cite{Engquist1981} and even the best ones exhibit poor quantization~\cite{Yacoby1996} and non-vanishing four-terminal resistance~\cite{dePicciotto2001}. 

Somewhat intermediate case is realized in quantum spin-Hall (QSH) insulators~\cite{Bernevig2006,Konig2007}. Here, the electric current is carried by a pair of 1D helical edge channels with opposite spin and chirality, thereby the backscattering is easier than in the quantum Hall case and more difficult compared to the conventional 1D case. In spite of the expected immunity to a non-magnetic disorder in a phase-coherent helical edge channel~\cite{ZhangRMP}, the mean-free path in experiments is relatively small and longer channels behave as quasi-classical diffusive conductors~\cite{Gusev2014,Tikhonov2015,Piatrusha2018,Wu2018}. The shorter, ballistic, channels exhibit four-terminal resistance which is poorly quantized and additive~\cite{Roth2009}. Often the resistance randomly drops below the quantum value $g_0^{-1}=h/e^2$ in local measurements~\cite{Konig2007,Li2017,Bendias2018} and below the expected fraction of $g_0^{-1}$ in non-local measurements~\cite{Olshanetsky2015}. This indicates that the measured signal is not the 1D conductance and is influenced by contact effects. Backscattering of helical electrons at a contact can be revealed in transport~\cite{Mani2016,ManiPRA2016} and noise measurements~\cite{Mani2017} as well as in spin injection~\cite{Aseev2013} and photogalvanic~\cite{Entin2016} experiments.

In this work, we elaborate the role played by the leads and ohmic contacts in resistance measurements in ballistic helical edge channels. A simple model of a phase-incoherent transport taking spin relaxation in the leads and contacts into account is presented for a realistic experimental setup. We observe that the four-terminal resistance is always below $g_0^{-1}$ and vanishes in the absence of spin relaxation. Similarly, a non-additivity of the edge resistances is observable in a two-terminal measurement. We bridge our results with the model of disordered contacts~\cite{Mani2016,ManiPRA2016,Mani2017} and estimate the spin relaxation resistance contribution in HgTe-based QSH devices. 

\begin{figure*}[t]
\begin{center}
\includegraphics[width=2\columnwidth]{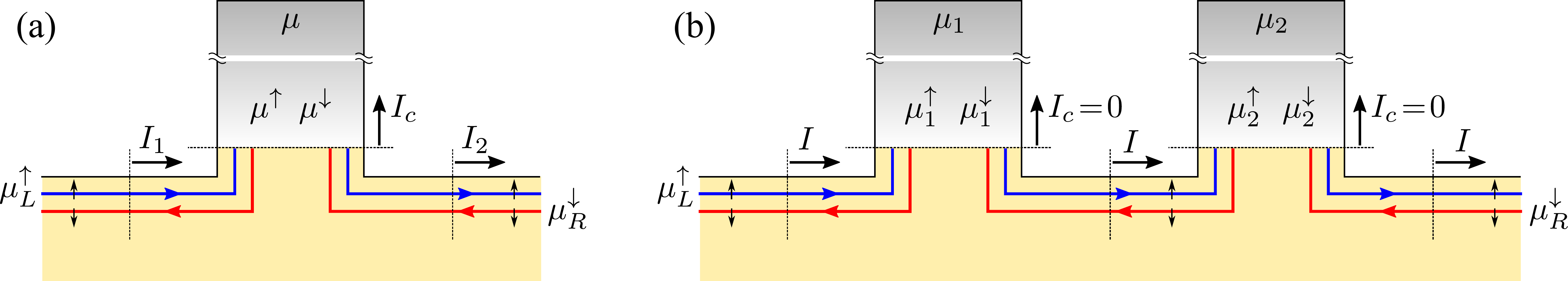}
\end{center}
\caption{(a) -- Currents and chemical potentials nearby a current or voltage lead coupled to the helical edge. The chemical potentials of the two incoming edge channels are set by the neighboring leads and denoted by $\mu_L^\uparrow$ and $\mu_R^\downarrow$. The chemical potential of the ohmic contact, measurable by an external voltage probe, is denoted by $\mu$. The chemical potentials of the two outgoing edge channels are set by those of the spin-up ($\mu^\uparrow$) and spin-down ($\mu^\downarrow$) electrons in the lead nearby the edge channel entrance/exit point.
The edge currents $I_1,\,I_2$ and the current in the ohmic contact $I_C$ are measured in the corresponding cross-sections marked by the dotted lines. (b) -- Currents and chemical potentials corresponding to a four-terminal resistance measurements of a helical edge channels.}\label{fig1}
\end{figure*}

Below we develop an experimentally relevant model of a multi-terminal bar for QSH measurements. In a typical experimental setup~\cite{Konig2007,Olshanetsky2015,Li2017}, a lithographic gate covers the inner part of the mesa, excluding the leads. All the leads are assumed identical and are represented by regions of two-dimensional electron gas. The leads have finite resistance and interconnect helical edge channels with the ohmic contacts. The ohmic contacts have negligible resistance and serve as macroscopic equilibrium reservoirs, connecting the device to external electric circuit. A basic element of such multiterminal structure consists of one lead and two pairs of incoming and outgoing helical edge channels, as depicted in Fig.~\ref{fig1}a. In the following we will consider the idealized case of ballistic topologically protected edge states, such that the spin relaxation occurs only in the leads and the ohmic contacts. In addition, we assume that the edge channels are perfectly coupled to the leads. This means that the chemical potentials of the outgoing edge channels coincide with those of the same-spin electrons in the leads nearby the bulk-edge transition point ($\mu^\downarrow$, $\mu^\uparrow$ in Fig.~\ref{fig1}a). All the leads are assumed to be quasi-1D, such that any dependence of the chemical potentials within the cross-section of the lead is neglected.

In situation of Fig.~\ref{fig1}a, the electric currents carried by the edge channels and the respective chemical potentials are related via a balance of electric currents:
\begin{align}
	I_1   = & I_c +I_2 \nonumber\\ 
	eI_1  = g_0\left(\mu_L^\uparrow-\mu^\downarrow\right);&\,\, eI_2 = g_0\left(\mu^\uparrow-\mu_R^\downarrow\right) \label{array1}\\ 
	eI_c  =g_L & \left(\frac{\mu^\uparrow+\mu^\downarrow}{2}-\mu\right), \nonumber
\end{align}
where the edge currents $I_1,\,I_2$ and the current in the contact $I_c$ are measured in the cross-sections marked by the dotted lines in Fig.~\ref{fig1}a and $g_L$ is the lead conductance. Note that the currents $I_1,\,I_2$ carried by individual edge channels are straightforward accessible experimentally, see, e.g.,~\cite{Couedo2016,Fei2017}. The chemical potential of the ohmic contact $\mu$ is the same for both spins. The spin relaxation in the leads enters via the balance of spin currents:
\begin{align}
	g_0\left(\mu_L^\uparrow-\mu^\uparrow\right)& =\frac{g_L}{2}\left(\mu^\uparrow-\mu\right)+eI_{\mathrm s}\nonumber\\
				g_0\left(\mu_R^\downarrow-\mu^\downarrow\right) & = \frac{g_L}{2}\left(\mu^\downarrow-\mu\right)-eI_{\mathrm s}\label{array2}\\
				eI_{\mathrm s} & =g_{\mathrm s}\left( \mu^\uparrow-\mu^\downarrow\right),\nonumber
		\end{align}
where $I_{\mathrm s}$ and $g_{\mathrm s}$ are, respectively, the spin relaxation current and the corresponding spin relaxation conductance in the lead, related via the last line in (\ref{array2}). Apart from the $I_{\mathrm s}$, the first two lines in (\ref{array2}) also contain the spin currents flowing in the ohmic contact. These contributions are given by the products of the corresponding chemical potential differences and the lead conductance per spin $g_L/2$. The solution of equations~(\ref{array1}) and~(\ref{array2}) for given  $I_1,\,I_2$ and $\mu$ is given by:
\begin{align}
\mu^\uparrow- \mu^\downarrow  = & \frac{e\left(I_1+I_2\right)}{2\left(G_{\mathrm s}+g_0\right)};\,\frac{\mu^\uparrow+\mu^\downarrow}{2}  = \mu+\frac{e\left(I_1-I_2\right)}{g_L} \nonumber\\
\mu^\downarrow-\mu_R^\downarrow & = \frac{e\left(I_2-I_1\right)}{2g_0}-\frac{e\left(I_1+I_2\right)}{2g_0}\frac{G_{\mathrm s}}{G_{\mathrm s}+g_0} \label{array3}\\
\mu_L^\uparrow-\mu^\uparrow & = \frac{e\left(I_1-I_2\right)}{2g_0}-\frac{e\left(I_1+I_2\right)}{2g_0}\frac{G_{\mathrm s}}{G_{\mathrm s}+g_0},\nonumber
		\end{align}
where we denoted a total spin relaxation conductance as $G_{\mathrm s}=g_{\mathrm s}+g_L/4$. Note that the term $g_L/4$ stands here for the relaxation via diffusion to the ohmic contacts, which remains effective even if the direct spin relaxation in the leads is suppressed ($g_{\mathrm s}=0$). As follows from the first line in (\ref{array3}), the chemical potentials $\mu^\uparrow$ and $\mu^\downarrow$ in the leads next to the QSH region differ unless the currents in the two edge channels are exactly opposite $I_1=-I_2$ (and vice-versa). This property was utilized in Ref.~\cite{Bruene2012} for the detection of the QSH effect via the inverse spin-Hall effect in the leads. Note also, that the last two lines in (\ref{array3}) imply that in the absence of spin relaxation ($G_{\mathrm s}=0$) the chemical potentials of both spin-up and spin-down electrons are insensitive to the presence of the current unbiased lead ($I_c=0$). It is this property that is responsible to a non-additivity of the helical edge resistances in the absence of spin relaxation.

Using equations (\ref{array3}) it is straightforward to derive the four-terminal resistance of the helical edge channel. The corresponding setup consists of the two sections of Fig.~\ref{fig1}a connected in series, as depicted in Fig.~\ref{fig1}b. A notable distinction of a four-terminal measurement is that no current flows in contacts 1 and 2 ($I_c=0$), which are used for voltage measurement, thus the currents in edge channels are all the same ($I$). Using an obvious relation $\mu_1^\uparrow-\mu_2^\downarrow=eI/g_0$ and equations analogous to the upper two lines in~(\ref{array3}), we obtain for the measured four-terminal resistance: 
\begin{equation}
			R_{4T}=\frac{\mu_1-\mu_2}{eI} = \frac{1}{g_0}\left[\frac{G_{\mathrm s}}{G_{\mathrm s}+g_0}\right]\label{4T}
\end{equation}

Eq.~(\ref{4T}) dictates that the measured $R_{4T}$ is always smaller than the quantum resistance $g_0^{-1}$. In the limit of large lead resistance and negligible spin relaxation ($G_{\mathrm s}\rightarrow0$) the $R_{4T}$ is suppressed  down to zero, similar to zeroing of the longitudinal resistance in the ordinary quantum Hall regime. Thus, apparently, the requirement of quantum phase-coherence in the leads raised in Refs.~\cite{Roth2009,ZhangRMP} is excessive as long as the spin relaxation is suppressed. 

In the opposite limit $G_{\mathrm s}\gg g_0$ the spin relaxation is strong and we obtain a small correction to the resistance quantum $R_{4T}\approx g_0^{-1}-G_{\mathrm s}^{-1}$. It is interesting to compare with Ref.~\cite{Mani2016}, which addresses a QSH measurement with disordered  current/voltage probes. Relevant to our case, that calculation predicts $R_{4T}=g_0^{-1}\left(1-D\right)/\left(1+D\right)$, where $D$ is a reflection probability at a contact (formula (17) with all $D$ the same~\cite{Mani2016}). This coincides with the result (\ref{4T}) given $D=\left(1+2G_{\mathrm s}/g_0\right)^{-1}$, thereby bridging the model of disordered probes~\cite{Mani2016,ManiPRA2016,Mani2017} with the spin relaxation in the leads in typical experiments.

\begin{figure}[t]
\begin{center}
\vspace{0mm}
\textsc{\textsc}\includegraphics[width=1\columnwidth]{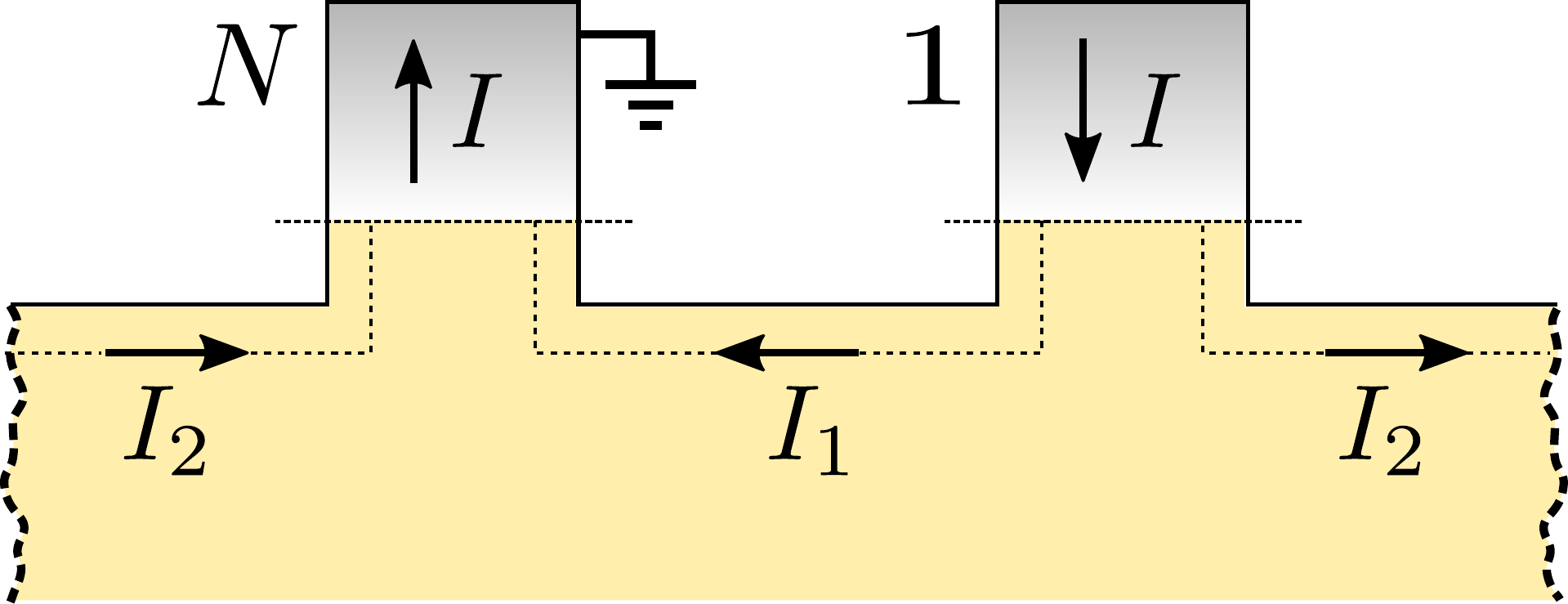}
\end{center}
\caption{ A sketch of a two-terminal measurement of a helical edge channel between the two neighboring leads of $N$-terminal bar.}\label{fig2}
\end{figure}

Next we calculate a two-terminal resistance of the ballistic helical edge channel. Such a measurement is configured in an $N$-terminal bar as depicted in Fig.~\ref{fig2}. The neighboring terminals 1 and $N$ serve, respectively, as the source and the drain for the total current $I$. One part of this current ($I_1$) flows counter-clockwise directly from 1 to $N$, while the other part ($I_2=I-I_1$) flows clockwise around the bar passing by the terminals 2,...,$N-1$ (not shown). Solving a set of equations analogous to (\ref{array3}) we find for currents $I/I_2=N$ and for the two-terminal resistance $R_{2T}\equiv\left(\mu_1-\mu_N\right)/eI$:

\begin{equation}
			R_{2T}= \frac{2}{g_L}+\frac{1}{2g_0}+\frac{\left(N-2\right)\left(G_{\mathrm s}+2g_0\right)}{\left(NG_{\mathrm s}+2g_0\right)\left(G_{\mathrm s}+g_0\right)}\frac{G_{\mathrm s}}{2g_0}\label{2T}
\end{equation}

The first two terms in (\ref{2T}) are, respectively, the inevitable contribution of the lead resistance and the resistance of two helical edges in parallel. The last term takes a non-additivity of the helical edge resistances into account, given the spin relaxation is finite. In the limit $G_{\mathrm s}\rightarrow0$, as well as for $N=2$, this term vanishes and we recover a result equivalent to the ordinary (spin-degenerate) quantum Hall effect. In the opposite limit $G_{\mathrm s}\gg g_0$ the edge resistances become completely additive and $R_{2T}$ is a sum of the lead resistance and the edge resistances $g_0^{-1}$ and $(N-1)g_0^{-1}$ connected in  parallel. In a multi-terminal bar with $N\rightarrow\infty$ we have $R_{2T}\approx2g_L^{-1}+g_0^{-1}+(2G_{\mathrm s})^{-1}$, i.e. the first-order correction here is opposite in sign compared to the $R_{4T}$ case.

Finally, we estimate the contribution of the spin relaxation resistance $G_{\mathrm s}^{-1}$ in experiments. In a typical QSH device the leads are formed by a two-dimensional electron gas with strong spin-orbit coupling, which results in a finite spin relaxation length $l_{\mathrm s}=\sqrt{D\tau_{\mathrm s}}$, where $D$ is the diffusion coefficient and $\tau_{\mathrm s}$ is the spin relaxation time. Consider a rectangular shaped lead of a width much smaller than the length in the direction of the ohmic contact, $L\gg w$. Relevant for experiments is the case of strong relaxation $L\gg l_{\mathrm s}$. Hence, $G_{\mathrm s}^{-1}\approx g_{\mathrm s}^{-1}\sim\left(l_{\mathrm s}/w\right)\rho$ provided $l_{\mathrm s}\gg w$, where $\rho\sim100\,{\rm \Omega}$ is the sheet resistivity in the leads. In HgTe quantum wells, the measurements of the weak anti-localization~\cite{Minkov2012,Kozlov2013} indicate very strong spin relaxation with $l_{\mathrm s}\sim w\sim1\,\mu$m. In this case, $G_{\mathrm s}^{-1}\sim\rho$, such that the expected contribution of the spin relaxation resistance to $R_{4T}$ and $R_{2T}$ is within a few percent. This estimate is consistent with the rule of thumb that the edge resistances are additive in the experiments, as well as with 
numerous observations of $R_{4T}$ below the quantum value $g_0^{-1}=h/e^2$ in local measurements~\cite{Konig2007,Bendias2018} and below the expected fraction of $g_0^{-1}$ in non-local measurements~\cite{Olshanetsky2015}. 

In summary, we have shown how the spin relaxation in the current/voltage leads affects the resistance measurements of ballistic QSH helical edge channels in experimentally relevant geometries. Negligible relaxation results in a vanishing four-terminal resistance and non-additive edge resistances in a two-terminal setup even if the quantum phase-coherence is not preserved, similar to the case of ordinary quantum Hall effect. Available experiments are in the opposite limit of strong spin relaxation, which explains a poor quality of the resistance quantization as well as the edge resistances smaller than expected ballistic value. 

We acknowledge  valuable discussions with K.E. Nagaev, S.A. Tarasenko, Z.D. Kvon, S.U. Piatrusha and M.L. Savchenko. This work was supported by the Russian Science Foundation project Nr. 16-42-01050.





%

%

\end{document}